\documentclass[aps,pra,reprint,amsmath,superscriptaddress]{revtex4-2}
\usepackage{graphicx}
\usepackage{amsmath}
\usepackage{siunitx}
\usepackage{hyperref}
\usepackage[nameinlink,capitalize]{cleveref}
\usepackage{braket}

\newcommand{\FIGWIDTHTC}{17.2cm}

\begin{document}
\title{Simulation of positronium laser cooling using the Lindblad master equation}
\date{\today}
\author{Kenji Shu}
\affiliation{RIKEN Center for Emergent Matter Science~(CEMS), 2-1 Hirosawa, Wako, Saitama 351-0198, Japan}
\affiliation{Photon Science Center, School of Engineering, The University of Tokyo, 2-11-16 Yayoi, Bunkyo-ku, Tokyo 113-0032, Japan}
\author{Sohma Shiraishi}
\affiliation{Department of Applied Physics, School of Engineering, The University of Tokyo, 7-3-1 Hongo, Bunkyo-ku, Tokyo 113-8656, Japan}
\author{Kosuke Yoshioka}
\email{yoshioka@fs.t.u-tokyo.ac.jp}
\affiliation{Photon Science Center, School of Engineering, The University of Tokyo, 2-11-16 Yayoi, Bunkyo-ku, Tokyo 113-0032, Japan}
\affiliation{Department of Applied Physics, School of Engineering, The University of Tokyo, 7-3-1 Hongo, Bunkyo-ku, Tokyo 113-8656, Japan}

\begin{abstract}

We present a formulation and numerical results for positronium~(Ps) laser cooling.
The formulation is based on the Lindblad master equation and follows the time evolution of the density matrix of Ps atoms.
It therefore accounts for atomic coherence, which is necessary to describe the interaction of Ps with the train of short laser pulses generated by the system developed by Shu \textit{et al.}~[K. Shu \textit{et al.}, Phys. Rev. A \textbf{109}, 043520 (2024)].
Using this formulation, we calculate the time evolution of the populations in each internal and momentum state and thereby quantitatively predict the momentum distribution after laser cooling.
We present the representative time evolution of the internal-state populations and momentum distribution, together with a comprehensive scan of the laser parameters used to optimize the cooling efficiency.
A prominent feature of the simulated distributions is sub-recoil cooling through velocity-selective coherent population trapping, a coherent effect captured by the quantum-mechanical treatment.

\end{abstract}

\maketitle

\section{Introduction}

Positronium, the bound state of an electron and a positron, is particularly well suited to precision tests of fundamental physics because it consists solely of two light leptons.
Its properties, including its energy levels, can be predicted precisely by quantum electrodynamics~(QED)\,\cite{karshenboim_precision_2005,adkins_precision_2022}.
Comparisons of these predictions with precision measurements provide rigorous tests of QED\,\cite{adkins_precision_2022,cassidy_experimental_2018,fee_measurement_1993,ishida_new_2014,sheldon_precision_2023}.
A significant discrepancy between theory and experiment could signal physics beyond the Standard Model\,\cite{adkins_precision_2022,frugiuele_current_2019}.
At present, the precision of Ps measurements remains lower than that of the corresponding theoretical predictions\,\cite{fee_measurement_1993,ishida_new_2014,sheldon_precision_2023}.
This limitation is particularly important in Ps spectroscopy, which offers high relative precision but is hindered by the high temperatures of the Ps gases used in experiments.
In the absence of an efficient cooling technique, previous precision measurements used Ps gases at several hundred kelvin.
The small mass of Ps amplifies the effects of these temperatures because both the mean speed and the velocity spread scale inversely with the square root of the mass.
Consequently, Doppler shifts and broadening, transit-time broadening, and the motional Stark effect can substantially degrade measurement precision and accuracy.
Two independent experiments have recently demonstrated laser cooling of Ps, including our own experiment\,\cite{shu_cooling_2024,gloggler_positronium_2024}.
Our cooling scheme produced an ultracold component whose velocity-distribution envelope corresponded to a temperature of approximately 1\,K\,\cite{shu_cooling_2024}.
Measurements using Ps at such temperatures could reduce velocity-related systematic effects and uncertainties, potentially enabling tests of bound-state QED at or beyond the current theoretical precision.

In the pursuit of ultracold temperatures, simulations of laser cooling are invaluable for identifying optimal cooling conditions and interpreting observed cooling effects through comparison with experiment\,\cite{liang_laser_1988,kumita_study_2002,shu_study_2016,zimmer_positronium_2021}.
These benefits are particularly important because experiments with Ps are challenging.
Its small mass and short lifetime require cooling lasers with unusual temporal and spectral properties, often necessitating dedicated laser development.
Simulations can therefore help specify the parameters required for laser design.
They can also guide experimental searches by identifying the relevant regions of parameter space.
Repeated experimental evaluations of the Ps cooling efficiency are impractical because the available Ps flux is limited by the scarcity of positrons.
Typical fluxes of slow positrons, which are required to produce Ps, range from $10^4$ to $10^7\,\mathrm{s^{-1}}$ and are substantially lower than typical atomic fluxes.
Moreover, access to high-intensity beams often requires an accelerator facility and is therefore limited; this was the case for our laser-cooling demonstration.
Accurate simulations can make efficient use of limited experimental time and thereby accelerate Ps cooling research.

Several analytical and numerical studies of Ps laser cooling have been reported\,\cite{liang_laser_1988,kumita_study_2002,shu_study_2016,zimmer_positronium_2021}.
Previous simulations intended for realistic applications have predominantly used rate equations to describe the Ps--laser interaction, thereby neglecting atomic coherence.
This approximation is valid for broadband-laser cooling when the field-intensity envelope varies slowly compared with the decoherence time of the relevant transition, typically the $1^3S_1$--$2^3P_J$ transition in Ps.
Recently, we developed a laser for Ps chirp cooling that emits a train of pulses with a pulse spacing comparable to the decoherence time, whereas each pulse is much shorter than that time\,\cite{yamada_theoretical_2021,shu_development_2024}.
These temporal characteristics induce coherent atomic transitions and violate the assumptions underlying the rate-equation approximation, necessitating a quantum-mechanical treatment of the atomic dynamics.
In this paper, we present simulations of the Ps chirp cooling process based on the Lindblad master equation.
We use this formalism to track the joint evolution of the Ps internal state and translational momentum distribution, including the non-negligible coherence induced by the pulse train.
We present representative numerical results and systematically investigate the dependence of the cooling efficiency on the laser parameters.
We also discuss possible extensions of the simulation.

\section{Formalism}

\subsection{Time evolution of positronium}

We use the Lindblad master equation to simulate Ps laser cooling.
This equation provides a quantum-mechanical treatment of the atoms and is required here because the train of short laser pulses induces atomic coherence\,\cite{meystre2007elements,ilinova_doppler_2011}.
The joint distribution over the internal states and translational momenta is encoded in the density matrix~$\rho(t)$, whose evolution is governed by
\begin{equation}
    \frac{\mathrm{d}}{\mathrm{d}t} \rho(t) = \frac{1}{i \hbar} \left[ H, \rho(t) \right] + L(\rho(t)),
\end{equation}
where $\hbar$ is the reduced Planck constant, $H$ is the Hamiltonian, and $L$ accounts for dissipation.
We retain the Ps--photon interaction through electric-dipole order.
The Hamiltonian~$H$ includes the Ps--laser interaction, with the laser fields represented as coherent states of the photon field.
Because the occupation numbers of all relevant laser modes are large, we neglect changes in these numbers due to absorption and stimulated emission.
Spontaneous emission and decay into $\gamma$ rays through self-annihilation are included in the dissipative term.

We consider one-dimensional cooling in the same geometry as our laser-cooling experiment.
The basis of the Hilbert space consists of eigenstates of the unperturbed Ps Hamiltonian.
We denote these states by $\ket{nlsJMk_\mathrm{p}}$.
The first five symbols specify the internal state: the principal quantum number~$n$, orbital angular momentum~$l$, total electron--positron spin~$s$, total angular momentum~$J$, and its projection~$M$ onto the quantization axis.
The signed integer~$k_\mathrm{p}$ labels the one-dimensional translational momentum $k_\mathrm{p} \Delta_\mathrm{p}$, where $\Delta_\mathrm{p}$ is the momentum-grid spacing.
The corresponding eigenenergy is
\begin{equation}
    E_{nlsJMk_\mathrm{p}} = E_{nlsJ}^\mathrm{b} + \frac{(k_\mathrm{p} \Delta_\mathrm{p})^2}{2 m_\mathrm{Ps}}.
\end{equation}
The first term is the binding energy, and the second is the kinetic energy of the center-of-mass motion.
The Ps mass~$m_\mathrm{Ps}$ is the sum of the electron and positron rest masses.
We restrict the quantum numbers to $s=1$ and $n=1,2$ to simulate Lyman--$\alpha$ laser cooling of \textit{ortho}-Ps, whose ground state is relatively long-lived.
The lifetime of ground-state \textit{para}-Ps, for which $s=0$, is sufficiently short that we neglect its initial population.
We also neglect mixing between the $s=0$ and $s=1$ states because no static magnetic field is considered.
The basis therefore contains three $1^3S_1$ states and nine $2^3P_J$ states with $J=0,\,1,$ and~2.
We require the first-order Doppler shift associated with one momentum step, $\omega_{0}\Delta_\mathrm{p} / (m_\mathrm{Ps} c)$, to be sufficiently smaller than the natural linewidth of the transition, where $\omega_0$ is the resonant angular frequency and $c$ is the speed of light.
This condition makes the Doppler-broadened absorption spectrum effectively smooth and renders momentum-discretization errors negligible.
We choose $\Delta_\mathrm{p} \simeq 8\,\mathrm{meV}\,c^{-1}$, for which the corresponding first-order Doppler shift is approximately $2 \pi \times \SI{10}{\MHz}$, smaller than the natural linewidth of approximately $2\pi \times \SI{50}{\MHz}$; both values are expressed as angular frequencies.
For Ps produced at several hundred kelvin by conventional methods, the momentum range addressed by laser cooling spans first-order Doppler shifts of order 100\,GHz.
The resulting momentum space is large because the low mass of Ps produces a broad momentum distribution.
We therefore performed parallel calculations on the supercomputer Fugaku at the RIKEN Center for Computational Science.
In our simulation code, the matrices in the Lindblad master equation are represented as sparse matrices.
A typical calculation reported here required a few days on tens of Fugaku compute nodes.
We also choose $\Delta_\mathrm{p}$ such that the resonant photon momentum~($\simeq 5.1\,\mathrm{eV}\,c^{-1}$) is an integer multiple of $\Delta_\mathrm{p}$.
This choice reduces the momentum-conservation errors introduced by discretizing the recoil from photon absorption and emission.
 
The electric-dipole interaction is described by the Hamiltonian $H_\mathrm{ed} = -\vec{d} \cdot \vec{E}$, where $\vec{d}$ is the electric-dipole moment and $\vec{E}$ is the electric field.
For the interaction between Ps and the laser fields, the matrix element of $H_{\mathrm{ed}}$ in the interaction picture is approximated as
\begin{widetext}
    \begin{equation}
	\bra{i} H_{\mathrm{ed}}^{\mathrm{I}}(t) \ket{j} = -\frac{1}{2} \vec{d_{ij}} \cdot \sum_{L} (\vec{E_L}(t) \exp(i \delta_L^- t) \tilde{\delta}_{P_i - \hbar k_\mathrm{c}^{L}, P_j} + \vec{E_L^\star}(t) \exp(i \delta_L^+ t) \tilde{\delta}_{P_i + \hbar k_\mathrm{c}^L, P_j}).
	\label{eq:e1_interaction_hamiltonian}
    \end{equation}
\end{widetext}
Here $i$ and $j$ index the eigenstates, $\vec{d}_{ij}$ is the corresponding matrix element of $\vec{d}$, and $\vec{E}_L(t)$ is the electric-field envelope of the $L$-th laser beam.
We define the envelope such that the complex electric field of the $L$-th laser at the spacetime point~$(t,x)$ is $\vec{E}_L(t) \exp[i(k_\mathrm{c}^L x - \omega_\mathrm{c}^L t)]$.
The quantities $\omega_\mathrm{c}^L$ and $k_\mathrm{c}^L=\omega_\mathrm{c}^L/c$ are the carrier angular frequency and wavenumber, respectively.
The angular-frequency detunings for the transition under consideration are $\delta_L^\pm = (E_i - E_j \pm \hbar \omega_\mathrm{c}^{L} ) / \hbar$.
The term $\tilde{\delta}_{P, Q}$ enforces momentum conservation on the grid with spacing $\Delta_{\mathrm{p}}$ during photon absorption and emission:
\begin{equation}
    \tilde{\delta}_{P, Q} = \begin{cases}
	1& \text{if $|P-Q| \leq \Delta_{\mathrm{p}}/2$}, \\
	0& \text{if $|P-Q| > \Delta_{\mathrm{p}}/2$}.
    \end{cases}
\end{equation}
The first term in parentheses in \cref{eq:e1_interaction_hamiltonian} represents stimulated emission, and the second represents absorption.
The main approximation is to replace the momentum of every emitted or absorbed laser photon by the carrier momentum $\hbar k_{\mathrm{c}}^{L}$.
The relative photon-momentum spread of the laser is approximately $100\,\mathrm{GHz} / 1\,\mathrm{PHz}\simeq10^{-4}$; the corresponding absolute spread is much smaller than both $\Delta_{\mathrm{p}}$ and the desired resolution of the simulated momentum distributions.
Under this approximation, the laser spectrum enters through the temporal field envelope.

The dissipative term~$L(\rho(t))$ includes spontaneous emission and self-annihilation:
\begin{widetext}
    \begin{equation}
	\bra{i} L(\rho(t)) \ket{j} = \left( -\frac{1}{2} \sum_{a,b} (\Gamma_{ab}^{\mathrm{sp.}} + \Gamma_{a}^{\mathrm{ann.}})(\delta_{ia}\rho_{aj}(t) + \delta_{ja}\rho_{ia}(t)) + \sum_{a,b} \Gamma_{ab}^{\mathrm{sp.}} \delta_{ib} \delta_{jb} \rho_{aa}(t) \right).
    \end{equation}
\end{widetext}
The indices $a$ and $b$ are summed over all eigenstates.
$\Gamma_{ab}^{\mathrm{sp.}}$ is the spontaneous-emission rate from state~$a$ to state~$b$, and $\Gamma_{a}^{\mathrm{ann.}}$ is the self-annihilation rate of state~$a$.
In this expression, we neglect the transfer of coherence from the excited states~($n=2$) to the ground states~($n=1$) through spontaneous emission.
This approximation reduces the number of nonzero elements that require costly evaluation and is valid provided that the laser does not create excited states that are coherent superpositions of different $M$ levels; such superpositions would transfer coherence to the ground states through spontaneous emission\,\cite{aspect_laser_1989}.
This condition is satisfied for a linearly polarized laser field that propagates perpendicular to the quantization axis and drives only $\pi$ transitions.
We therefore choose the quantization axis of the Ps internal states to be perpendicular to the laser propagation direction and parallel to the polarization direction.
Consequently, each excited state is coupled by the lasers to only one ground-state Zeeman sublevel.
We approximate the spontaneous-emission rates between eigenstates by integrating the momentum-resolved rate over the range of momentum differences represented by each discrete final state:
\begin{widetext}
    \begin{equation}
	\Gamma_{ab}^{\mathrm{sp.}} = \frac{\alpha}{2\pi c^2} \int_{-\infty}^{\infty} \mathrm{d} \omega \, \omega^3 \delta \left( \omega \left(1 + \frac{\hbar \omega}{2 m_\mathrm{Ps} c^2} \right) - \omega_{ba} \right) \int_{z^-(\omega)}^{z^+(\omega)} \mathrm{d} z \, \int_{0}^{2\pi} \mathrm{d} \phi \, \sum_{\lambda} |\vec{d}_{ba} \cdot \vec{\varepsilon}_{z,\phi,\lambda}|^2.
    \end{equation}
\end{widetext}
Here $\alpha$ is the fine-structure constant, and $\omega_{ba}$ is the angular-frequency difference between states~$a$ and~$b$.
$\vec{\varepsilon}_{z,\phi,\lambda}$\,($\lambda=1,\,2$) is a photon polarization vector for emission at polar angle $\arccos(z)$ and azimuthal angle~$\phi$ relative to the Ps momentum axis.
The integration range of $z$ is chosen such that the projection of the photon momentum $\hbar \omega / c$ matches the momentum difference between the two states.
The integration limits are therefore $z^\pm (\omega) = \left( k_\mathrm{p,b} - k_\mathrm{p,a} \pm \frac{1}{2} \right) \frac{c \Delta_\mathrm{p}}{\hbar \omega}$.
The terms in the Dirac delta function account for the recoil shift.
We neglect the Doppler correction to the recoil because it changes the recoil momentum by less than 0.1\%.
This correction is negligible when the resulting momentum distribution is much broader than the associated momentum correction.
For $\Gamma_{a}^{\mathrm{ann.}}$, we use measured or calculated decay rates for the internal states under consideration.

\subsection{Laser field}

We consider chirp cooling in which Ps is irradiated by trains of laser pulses whose center frequencies chirp upward.
We model the electric-field envelope produced by the Chirped Pulse-Train Generator~(CPTG)\,\cite{yamada_theoretical_2021,shu_cooling_2024} as
\begin{widetext}
    \begin{equation}
	\label{eq:laser_field}
	E_{\mathrm{f}}^{\mathrm{CPT}}(t) = A \exp( -i m(t) n_\mathrm{c} \omega_\mathrm{r} t + i \beta \sin(\Omega t) )\sum_{n=-\infty}^{\infty} \exp \left( -\frac{|n| \omega_\mathrm{r}}{\sigma} \right) \exp (-i n \omega_\mathrm{r} t).
    \end{equation}
\end{widetext}
Here $A$ sets the laser intensity, $n_{\mathrm{c}}$ is an integer that determines the chirp rate, and $\omega_{\mathrm{r}} \simeq 2\pi \times 236$\,MHz is the pulse-repetition angular frequency.
The modulation depth $\beta \simeq 0.403$\,rad and angular frequency $\Omega \simeq 2\pi \times 78.8$\,MHz make the effective laser spectrum nearly continuous.
The summation index is $n$, and $\sigma$ determines the spectral width of a single pulse, which we refer to as the instantaneous spectral width.
We choose $\beta$ to equalize the intensities of the generated sidebands.
The function $m(t)$ returns the index of the pulse at time~$t$:
\begin{equation}
  m(t) = \left\lfloor \frac{\omega_\mathrm{r} t}{2 \pi} \right\rfloor.
\end{equation}
The summation in \cref{eq:laser_field} forms the envelope of each pulse, while the preceding exponential factors impose the pulse-to-pulse frequency up-chirp and spectral broadening.
The broadening prevents the effective laser spectrum from becoming too sparse.
We choose $\Omega$ to be a divisor of $\omega_\mathrm{r}$ and comparable to the natural linewidth of the transition.
In the Hamiltonian~$H_{\mathrm{ed}}^{\mathrm{I}}$, we use the envelope $E_L(t)=(E_{\mathrm{f}}^{\mathrm{CPT}}(t))^3$ because we assume third-harmonic conversion of the near-infrared CPTG output to the $1S$--$2P$ resonance wavelength of approximately 243\,nm.

\section{Results and Discussion}

\begin{table}
    \caption{Laser parameters used in the simulation to reproduce the laser developed for the Ps laser-cooling demonstration.}
    \begin{ruledtabular}
    \begin{tabular}{cr}
	Parameter name & Value \\
	\hline
	Chirp rate & 500\,GHz\,$\mu$s$^{-1}$ \\
	Instantaneous spectral width & 9\,GHz \\
	Average intensity & 1\,kW\,cm$^{-2}$ \\
	Pulse train duration & 100\,ns \\
	Polarization & linear
    \end{tabular}
    \end{ruledtabular}
    \label{tab:CPTG_base_parameter}
\end{table}
\begin{figure*}
    \includegraphics[width=\FIGWIDTHTC]{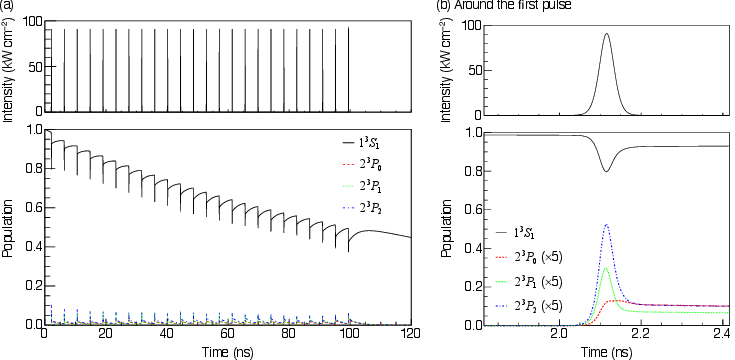}
    \caption{
	The top panels show the laser intensity, and the bottom panels show the populations of the internal states as functions of time.
	(a) Full time range through the end of the cooling-laser irradiation.
	(b) Expanded view around the arrival of the first laser pulse.
    }
    \label{fig:time_evolution_internal_states}
\end{figure*}
First, we present a simulation using the parameters of the developed laser.
The initial Ps ensemble is unpolarized, occupies the ground state, and has a Maxwell--Boltzmann momentum distribution at 300\,K.
We include two counterpropagating laser fields to model the one-dimensional irradiation geometry.
The parameters are summarized in \cref{tab:CPTG_base_parameter} and reproduce those of the laser used in the Ps laser-cooling demonstration.
After the frequency chirp, the carrier frequency of the final pulse is 1,233,592\,GHz, corresponding to a detuning of approximately $-7$\,GHz from the degeneracy-weighted resonance frequency of stationary Ps.
\Cref{fig:time_evolution_internal_states} shows the laser intensity and the populations of the internal states as functions of time.
Each laser pulse drives transitions from the $1S$ to the $2P$ states, after which the excited states decay exponentially to the ground states through spontaneous emission.
As shown in \cref{fig:time_evolution_internal_states}\,(b), the $2^3P_1$ and $2^3P_2$ populations peak during each laser pulse and then decrease as the pulse intensity falls.
This behavior indicates that each pulse drives the transition on a time scale much shorter than the excited-state lifetime and that its pulse area exceeds $\pi$, allowing the Rabi oscillation to proceed beyond a $\pi$ pulse.
The result demonstrates the need for the Lindblad master equation, which retains the atomic coherence required to describe Rabi oscillations.
The three transitions have different Rabi frequencies because their strengths depend on the matrix elements and degeneracies of the excited states.
In contrast, rate equations cannot describe these pulse-driven coherent transitions and instead predict excited-state populations that increase monotonically toward their steady-state ratios.
This limitation can produce non-negligible errors in the cooling efficiency because laser-induced transition probabilities govern the cooling process.

\begin{figure*}
    \includegraphics[width=\FIGWIDTHTC]{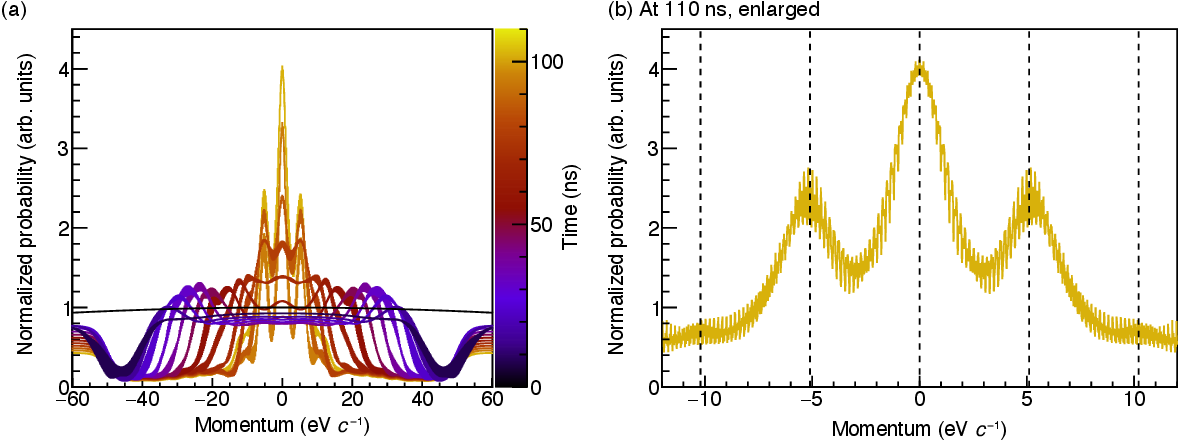}
    \caption{
	(a) Evolution of the ground-state momentum distribution.
	Distributions at selected times are superimposed as colored curves, with the times indicated by the color scale on the right.
	The probability density on the vertical axis is normalized to its peak value at 0\,ns.
	Curves are shown at 0\,ns, $1.5t_{\mathrm{period}}$, $2.5t_{\mathrm{period}}$, at subsequent intervals of $t_{\mathrm{period}}$ through the end of the cooling pulse train, and at 110\,ns after the excited states have decayed spontaneously into the $1S$ manifold.
	Here $t_{\mathrm{period}}$ is the pulse spacing within the train.
	(b) Momentum distribution at 110\,ns, with an expanded view around zero momentum. The vertical dashed lines mark integer multiples of the cooling-photon momentum, 5.1\,eV\,$c^{-1}$.
    }
    \label{fig:momentum_distribution}
\end{figure*}
We evaluate the cooling effect from the ground-state momentum distribution.
Its simulated evolution is shown in \cref{fig:momentum_distribution}\,(a).
Momentum is expressed in units of electronvolts divided by the speed of light.
The cooling-photon momentum is 5.1\,eV\,$c^{-1}$; a Ps atom with this momentum has a speed of $1.5\times 10^3$\,m\,s$^{-1}$.
The corresponding first-order Doppler shift for the 243-nm (1.23-PHz) cooling light is 6.2\,GHz.
Cooling appears as a depletion of fast Ps and an accumulation of slower Ps.
As the spectrum of each cooling pulse shifts upward toward the Ps $1S$--$2P$ resonance, the momentum range addressed by chirp cooling expands from high to low absolute momenta.
After the cooling laser sweeps over a momentum range of approximately 40\,eV\,$c^{-1}$ on each side of zero, three narrow peaks emerge.
The combined envelope of these peaks has a full width at half maximum of approximately 10\,eV\,$c^{-1}$.
This recoil-scale envelope indicates cooling close to the single-photon recoil limit.

A striking feature of the cooled momentum distribution is a series of narrow peaks centered at zero momentum and neighboring momenta separated by approximately 5.1\,eV\,$c^{-1}$, the single-photon recoil momentum.
The individual peak widths are approximately 3\,eV\,$c^{-1}$, substantially narrower than the recoil momentum.
We attribute this sub-recoil cooling to velocity-selective coherent population trapping\,\cite{aspect_laser_1989,papoff_transient_1992}.
Under coherent irradiation, superpositions of ground states with different momenta form eigenstates that do not couple to the excited states and are therefore dark states.
This decoupling results from destructive interference among the transition amplitudes from the constituent momentum states.
Repeated absorption and spontaneous-emission cycles selectively accumulate population in these dark states.
Their momentum widths are determined by the laser--atom interaction time and are already below the recoil momentum for the present laser configuration.
This sub-recoil effect may enable cooling of Ps to still lower temperatures.

\begin{figure*}
    \includegraphics[width=\FIGWIDTHTC]{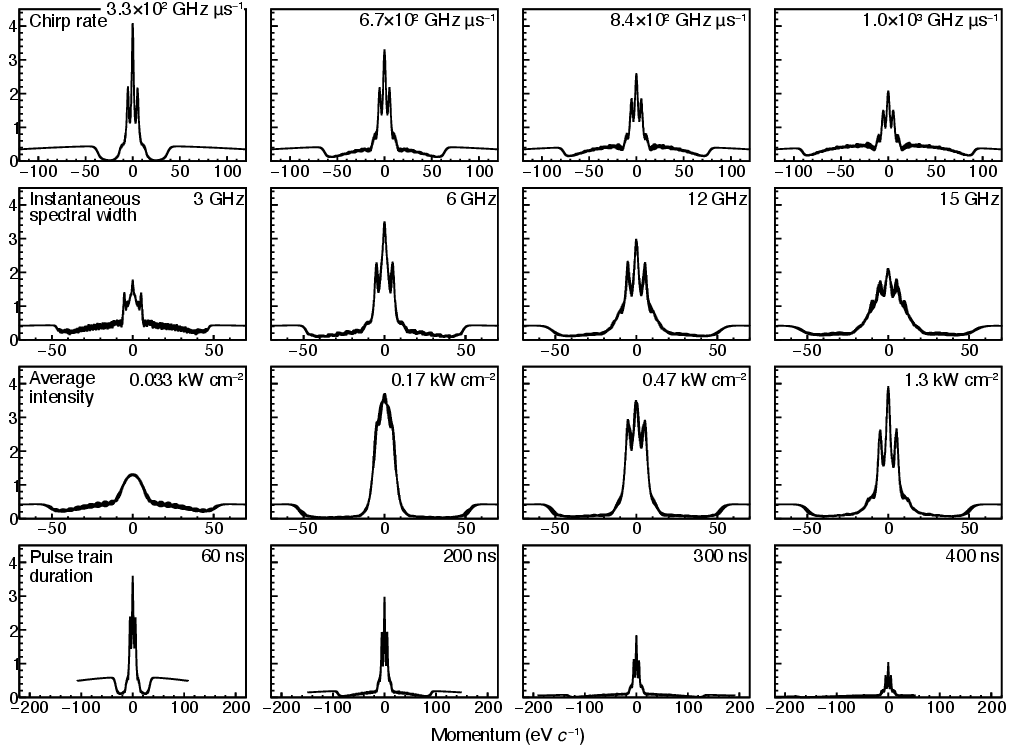}
    \caption{
	Momentum distributions for various laser parameters.
	In each row, the parameter named at the upper left of the leftmost panel is varied across the columns, and its value is shown at the upper right of each panel.
	All panels share a common vertical range and use the same probability normalization as \cref{fig:momentum_distribution}.
	The horizontal momentum range differs among rows but is the same for all columns within a given row.
    }
    \label{fig:parameter_space_scan}
\end{figure*}
We also vary the laser parameters to determine their effects on cooling efficiency and identify optimal values.
The four parameters in \cref{tab:CPTG_base_parameter} other than polarization are varied individually, with the remaining three fixed at their tabulated values.
The carrier frequency of the final pulse at the end of the frequency chirp is held at the value used in the simulations shown in \cref{fig:time_evolution_internal_states} and \cref{fig:momentum_distribution}, except when the instantaneous spectral width is varied.
For an instantaneous spectral width~$\Delta\nu_{\mathrm{inst}}$, this end-of-chirp carrier frequency is shifted from its value for the 9-GHz reference by $-(\Delta\nu_{\mathrm{inst}}-9\,\mathrm{GHz})/2$; for example, the shifts are $-1.5$\,GHz at 12\,GHz and $+1.5$\,GHz at 6\,GHz.
This adjustment keeps the high-frequency edge of the spectrum approximately fixed and suppresses the blue-detuned component that would otherwise cause laser heating at the end of cooling.
We also scale the laser intensity linearly with the width so that the spectral power density of each pulse is the same under all conditions.
The momentum distributions are evaluated 10\,ns after the end of the cooling pulse train.
The results are shown in \cref{fig:parameter_space_scan}.
These comparisons indicate that the parameters in \cref{tab:CPTG_base_parameter} are close to optimal.

The optimal chirp rate, instantaneous spectral width, and average intensity are qualitatively consistent with previous estimates based on the fundamental properties of Ps and the transition used for one-dimensional cooling\,\cite{shu_development_2024}.
The chirp rate should match the effective deceleration rate, which is determined by the photon recoil, spontaneous-emission rate, and excitation probability.
The instantaneous spectral width is optimal when it is comparable to both the recoil-associated Doppler shift and the fine-structure splitting of the transitions.
This choice balances efficient excitation without excessive spin polarization against the need for a sharp blue edge of the laser spectrum to avoid heating.
The intensity required for appreciable cooling is also consistent with the saturation intensity estimated for the present frequency-shifted broadband pulses.
At 0.17\,kW\,cm$^{-2}$, depletion of the fast Ps population over the swept range of approximately 15--40\,eV\,$c^{-1}$ is more efficient than at higher intensities.
This behavior likely occurs because the pulse energy is close to that required for coherent $\pi$-pulse excitation.
At higher intensities, the narrow peaks attributed to velocity-selective coherent population trapping become pronounced as the transition is fully saturated.

Longer cooling durations increase the range of velocities that can be cooled, but in one dimension the number of Ps atoms remaining near zero velocity is maximized at a duration of approximately 100\,ns.
The swept frequency range grows linearly with time, as does the addressed population if the velocity distribution is approximately flat over that range.
By contrast, the ground-state Ps population decreases exponentially through annihilation into $\gamma$ rays, with a lifetime of approximately 142\,ns.
The optimal cooling duration therefore depends on the intended application.
For studies of quantum degeneracy, the relevant figure of merit may be the number of stopped Ps atoms.
For precision spectroscopy, the optimum instead balances the loss of statistics against reductions in Doppler and transit-time broadening.

\section{Conclusion}

We developed a numerical simulation of one-dimensional Ps chirp cooling based on the Lindblad master equation that explicitly treats atomic coherence together with the internal and translational degrees of freedom.
The calculation resolved the coherent dynamics induced by a train of short laser pulses and predicted compression of an initial 300-K momentum distribution into a recoil-scale envelope.
Sub-recoil peaks were also obtained and attributed to velocity-selective coherent population trapping.
Scans of the laser parameters showed that the parameters of the developed cooling laser were close to optimal and clarified how each parameter affected the cooled momentum distribution.
The simulation platform developed here can also be applied to schemes that use coherent interactions driven by short laser pulses for more efficient cooling\,\cite{malamant_coherent_2024}.

While the present simulation provides accurate predictions for one-dimensional cooling, it does not track the trajectories of Ps atoms in real space, and it assumes that the Ps atoms always remain within the cooling laser field.
Extending the simulation to two- or three-dimensional chirp cooling, or evaluating the dynamics of the real-space distribution of Ps under cooling, is computationally challenging with the resources currently available.
We therefore consider a Monte Carlo model to be a practical approach for extending the calculation to higher-dimensional and real-space dynamics.
An empirical function describing the cooling dynamics, or the relevant parameters of the Monte Carlo model, could be calibrated against the predictions of the present one-dimensional simulation.
The calibrated model could then be used to predict higher-dimensional cooling and evaluate the spatial distribution.

\begin{acknowledgments}

The authors thank Mr. Takuto Kobayashi and Mr. Ryosuke Uozumi of the University of Tokyo for their contributions to developing and testing the simulation code and numerical model.
This work used computational resources of the supercomputer Fugaku provided by the RIKEN Center for Computational Science through the HPCI System Research Project (Project IDs: hp230215 and hp240397).
This work was supported by JSPS KAKENHI Grant Numbers JP24K00639 and JP24H00217, by the MEXT Quantum Leap Flagship Program (MEXT Q-LEAP), Grant Number JPMXS0118067246, and by the RIKEN TRIP initiative (HIKARI-COOL Tokyo).
This work was also supported by JST PRESTO Grant Number JPMJPR24F3, Japan.

\end{acknowledgments}


\begin{thebibliography}{20}%
\makeatletter
\providecommand \@ifxundefined [1]{%
 \@ifx{#1\undefined}
}%
\providecommand \@ifnum [1]{%
 \ifnum #1\expandafter \@firstoftwo
 \else \expandafter \@secondoftwo
 \fi
}%
\providecommand \@ifx [1]{%
 \ifx #1\expandafter \@firstoftwo
 \else \expandafter \@secondoftwo
 \fi
}%
\providecommand \natexlab [1]{#1}%
\providecommand \enquote  [1]{``#1''}%
\providecommand \bibnamefont  [1]{#1}%
\providecommand \bibfnamefont [1]{#1}%
\providecommand \citenamefont [1]{#1}%
\providecommand \href@noop [0]{\@secondoftwo}%
\providecommand \href [0]{\begingroup \@sanitize@url \@href}%
\providecommand \@href[1]{\@@startlink{#1}\@@href}%
\providecommand \@@href[1]{\endgroup#1\@@endlink}%
\providecommand \@sanitize@url [0]{\catcode `\\12\catcode `\$12\catcode
  `\&12\catcode `\#12\catcode `\^12\catcode `\_12\catcode `\%12\relax}%
\providecommand \@@startlink[1]{}%
\providecommand \@@endlink[0]{}%
\providecommand \url  [0]{\begingroup\@sanitize@url \@url }%
\providecommand \@url [1]{\endgroup\@href {#1}{\urlprefix }}%
\providecommand \urlprefix  [0]{URL }%
\providecommand \Eprint [0]{\href }%
\providecommand \doibase [0]{https://doi.org/}%
\providecommand \selectlanguage [0]{\@gobble}%
\providecommand \bibinfo  [0]{\@secondoftwo}%
\providecommand \bibfield  [0]{\@secondoftwo}%
\providecommand \translation [1]{[#1]}%
\providecommand \BibitemOpen [0]{}%
\providecommand \bibitemStop [0]{}%
\providecommand \bibitemNoStop [0]{.\EOS\space}%
\providecommand \EOS [0]{\spacefactor3000\relax}%
\providecommand \BibitemShut  [1]{\csname bibitem#1\endcsname}%
\let\auto@bib@innerbib\@empty
\bibitem [{\citenamefont {Karshenboim}(2005)}]{karshenboim_precision_2005}%
  \BibitemOpen
  \bibfield  {author} {\bibinfo {author} {\bibfnamefont {S.~G.}\ \bibnamefont
  {Karshenboim}},\ }\bibfield  {title} {\bibinfo {title} {Precision physics of
  simple atoms: {QED} tests, nuclear structure and fundamental constants},\
  }\href {https://doi.org/10.1016/j.physrep.2005.08.008} {\bibfield  {journal}
  {\bibinfo  {journal} {Phys. Rep.}\ }\textbf {\bibinfo {volume} {422}},\
  \bibinfo {pages} {1} (\bibinfo {year} {2005})}\BibitemShut {NoStop}%
\bibitem [{\citenamefont {Adkins}\ \emph {et~al.}(2022)\citenamefont {Adkins},
  \citenamefont {Cassidy},\ and\ \citenamefont
  {Pérez-Ríos}}]{adkins_precision_2022}%
  \BibitemOpen
  \bibfield  {author} {\bibinfo {author} {\bibfnamefont {G.~S.}\ \bibnamefont
  {Adkins}}, \bibinfo {author} {\bibfnamefont {D.~B.}\ \bibnamefont
  {Cassidy}},\ and\ \bibinfo {author} {\bibfnamefont {J.}~\bibnamefont
  {Pérez-Ríos}},\ }\bibfield  {title} {\bibinfo {title} {Precision
  spectroscopy of positronium: {Testing} bound-state {QED} theory and the
  search for physics beyond the {Standard} {Model}},\ }\href
  {https://doi.org/10.1016/j.physrep.2022.05.002} {\bibfield  {journal}
  {\bibinfo  {journal} {Phys. Rep.}\ }\textbf {\bibinfo {volume} {975}},\
  \bibinfo {pages} {1} (\bibinfo {year} {2022})}\BibitemShut {NoStop}%
\bibitem [{\citenamefont {Cassidy}(2018)}]{cassidy_experimental_2018}%
  \BibitemOpen
  \bibfield  {author} {\bibinfo {author} {\bibfnamefont {D.~B.}\ \bibnamefont
  {Cassidy}},\ }\bibfield  {title} {\bibinfo {title} {Experimental progress in
  positronium laser physics},\ }\href
  {https://doi.org/10.1140/epjd/e2018-80721-y} {\bibfield  {journal} {\bibinfo
  {journal} {Eur. Phys. J. D}\ }\textbf {\bibinfo {volume} {72}},\ \bibinfo
  {pages} {53} (\bibinfo {year} {2018})}\BibitemShut {NoStop}%
\bibitem [{\citenamefont {Fee}\ \emph {et~al.}(1993)\citenamefont {Fee},
  \citenamefont {Mills}, \citenamefont {Chu}, \citenamefont {Shaw},
  \citenamefont {Danzmann}, \citenamefont {Chichester},\ and\ \citenamefont
  {Zuckerman}}]{fee_measurement_1993}%
  \BibitemOpen
  \bibfield  {author} {\bibinfo {author} {\bibfnamefont {M.~S.}\ \bibnamefont
  {Fee}}, \bibinfo {author} {\bibfnamefont {A.~P.}\ \bibnamefont {Mills}},
  \bibinfo {author} {\bibfnamefont {S.}~\bibnamefont {Chu}}, \bibinfo {author}
  {\bibfnamefont {E.~D.}\ \bibnamefont {Shaw}}, \bibinfo {author}
  {\bibfnamefont {K.}~\bibnamefont {Danzmann}}, \bibinfo {author}
  {\bibfnamefont {R.~J.}\ \bibnamefont {Chichester}},\ and\ \bibinfo {author}
  {\bibfnamefont {D.~M.}\ \bibnamefont {Zuckerman}},\ }\bibfield  {title}
  {\bibinfo {title} {Measurement of the positronium {$1^3S_1$}-{$2^3S_1$}
  interval by continuous-wave two-photon excitation},\ }\href
  {https://doi.org/10.1103/PhysRevLett.70.1397} {\bibfield  {journal} {\bibinfo
   {journal} {Phys. Rev. Lett.}\ }\textbf {\bibinfo {volume} {70}},\ \bibinfo
  {pages} {1397} (\bibinfo {year} {1993})}\BibitemShut {NoStop}%
\bibitem [{\citenamefont {Ishida}\ \emph {et~al.}(2014)\citenamefont {Ishida},
  \citenamefont {Namba}, \citenamefont {Asai}, \citenamefont {Kobayashi},
  \citenamefont {Saito}, \citenamefont {Yoshida}, \citenamefont {Tanaka},\ and\
  \citenamefont {Yamamoto}}]{ishida_new_2014}%
  \BibitemOpen
  \bibfield  {author} {\bibinfo {author} {\bibfnamefont {A.}~\bibnamefont
  {Ishida}}, \bibinfo {author} {\bibfnamefont {T.}~\bibnamefont {Namba}},
  \bibinfo {author} {\bibfnamefont {S.}~\bibnamefont {Asai}}, \bibinfo {author}
  {\bibfnamefont {T.}~\bibnamefont {Kobayashi}}, \bibinfo {author}
  {\bibfnamefont {H.}~\bibnamefont {Saito}}, \bibinfo {author} {\bibfnamefont
  {M.}~\bibnamefont {Yoshida}}, \bibinfo {author} {\bibfnamefont
  {K.}~\bibnamefont {Tanaka}},\ and\ \bibinfo {author} {\bibfnamefont
  {A.}~\bibnamefont {Yamamoto}},\ }\bibfield  {title} {\bibinfo {title} {New
  precision measurement of hyperfine splitting of positronium},\ }\href
  {https://doi.org/10.1016/j.physletb.2014.05.083} {\bibfield  {journal}
  {\bibinfo  {journal} {Phys. Lett. B}\ }\textbf {\bibinfo {volume} {734}},\
  \bibinfo {pages} {338} (\bibinfo {year} {2014})}\BibitemShut {NoStop}%
\bibitem [{\citenamefont {Sheldon}\ \emph {et~al.}(2023)\citenamefont
  {Sheldon}, \citenamefont {Babij}, \citenamefont {Reeder}, \citenamefont
  {Hogan},\ and\ \citenamefont {Cassidy}}]{sheldon_precision_2023}%
  \BibitemOpen
  \bibfield  {author} {\bibinfo {author} {\bibfnamefont {R.}~\bibnamefont
  {Sheldon}}, \bibinfo {author} {\bibfnamefont {T.}~\bibnamefont {Babij}},
  \bibinfo {author} {\bibfnamefont {S.}~\bibnamefont {Reeder}}, \bibinfo
  {author} {\bibfnamefont {S.}~\bibnamefont {Hogan}},\ and\ \bibinfo {author}
  {\bibfnamefont {D.}~\bibnamefont {Cassidy}},\ }\bibfield  {title} {\bibinfo
  {title} {Precision {Microwave} {Spectroscopy} of the {Positronium}
  {2$^3$S$_1\to$2$^3$P$_2$} {Interval}},\ }\href
  {https://doi.org/10.1103/PhysRevLett.131.043001} {\bibfield  {journal}
  {\bibinfo  {journal} {Phys. Rev. Lett.}\ }\textbf {\bibinfo {volume} {131}},\
  \bibinfo {pages} {043001} (\bibinfo {year} {2023})}\BibitemShut {NoStop}%
\bibitem [{\citenamefont {Frugiuele}\ \emph {et~al.}(2019)\citenamefont
  {Frugiuele}, \citenamefont {Pérez-Ríos},\ and\ \citenamefont
  {Peset}}]{frugiuele_current_2019}%
  \BibitemOpen
  \bibfield  {author} {\bibinfo {author} {\bibfnamefont {C.}~\bibnamefont
  {Frugiuele}}, \bibinfo {author} {\bibfnamefont {J.}~\bibnamefont
  {Pérez-Ríos}},\ and\ \bibinfo {author} {\bibfnamefont {C.}~\bibnamefont
  {Peset}},\ }\bibfield  {title} {\bibinfo {title} {Current and future
  perspectives of positronium and muonium spectroscopy as dark sectors probe},\
  }\href {https://doi.org/10.1103/PhysRevD.100.015010} {\bibfield  {journal}
  {\bibinfo  {journal} {Phys. Rev. D}\ }\textbf {\bibinfo {volume} {100}},\
  \bibinfo {pages} {015010} (\bibinfo {year} {2019})}\BibitemShut {NoStop}%
\bibitem [{\citenamefont {Shu}\ \emph {et~al.}(2024{\natexlab{a}})\citenamefont
  {Shu}, \citenamefont {Tajima}, \citenamefont {Uozumi}, \citenamefont
  {Miyamoto}, \citenamefont {Shiraishi}, \citenamefont {Kobayashi},
  \citenamefont {Ishida}, \citenamefont {Yamada}, \citenamefont {Gladen},
  \citenamefont {Namba}, \citenamefont {Asai}, \citenamefont {Wada},
  \citenamefont {Mochizuki}, \citenamefont {Hyodo}, \citenamefont {Ito},
  \citenamefont {Michishio}, \citenamefont {O’Rourke}, \citenamefont
  {Oshima},\ and\ \citenamefont {Yoshioka}}]{shu_cooling_2024}%
  \BibitemOpen
  \bibfield  {author} {\bibinfo {author} {\bibfnamefont {K.}~\bibnamefont
  {Shu}}, \bibinfo {author} {\bibfnamefont {Y.}~\bibnamefont {Tajima}},
  \bibinfo {author} {\bibfnamefont {R.}~\bibnamefont {Uozumi}}, \bibinfo
  {author} {\bibfnamefont {N.}~\bibnamefont {Miyamoto}}, \bibinfo {author}
  {\bibfnamefont {S.}~\bibnamefont {Shiraishi}}, \bibinfo {author}
  {\bibfnamefont {T.}~\bibnamefont {Kobayashi}}, \bibinfo {author}
  {\bibfnamefont {A.}~\bibnamefont {Ishida}}, \bibinfo {author} {\bibfnamefont
  {K.}~\bibnamefont {Yamada}}, \bibinfo {author} {\bibfnamefont {R.~W.}\
  \bibnamefont {Gladen}}, \bibinfo {author} {\bibfnamefont {T.}~\bibnamefont
  {Namba}}, \bibinfo {author} {\bibfnamefont {S.}~\bibnamefont {Asai}},
  \bibinfo {author} {\bibfnamefont {K.}~\bibnamefont {Wada}}, \bibinfo {author}
  {\bibfnamefont {I.}~\bibnamefont {Mochizuki}}, \bibinfo {author}
  {\bibfnamefont {T.}~\bibnamefont {Hyodo}}, \bibinfo {author} {\bibfnamefont
  {K.}~\bibnamefont {Ito}}, \bibinfo {author} {\bibfnamefont {K.}~\bibnamefont
  {Michishio}}, \bibinfo {author} {\bibfnamefont {B.~E.}\ \bibnamefont
  {O’Rourke}}, \bibinfo {author} {\bibfnamefont {N.}~\bibnamefont {Oshima}},\
  and\ \bibinfo {author} {\bibfnamefont {K.}~\bibnamefont {Yoshioka}},\
  }\bibfield  {title} {\bibinfo {title} {Cooling positronium to ultralow
  velocities with a chirped laser pulse train},\ }\href
  {https://doi.org/10.1038/s41586-024-07912-0} {\bibfield  {journal} {\bibinfo
  {journal} {Nature}\ }\textbf {\bibinfo {volume} {633}},\ \bibinfo {pages}
  {793} (\bibinfo {year} {2024}{\natexlab{a}})}\BibitemShut {NoStop}%
\bibitem [{\citenamefont {Glöggler}\ \emph {et~al.}(2024)\citenamefont
  {Glöggler}, \citenamefont {Gusakova}, \citenamefont {Rienäcker},
  \citenamefont {Camper}, \citenamefont {Caravita}, \citenamefont {Huck},
  \citenamefont {Volponi}, \citenamefont {Wolz}, \citenamefont {Penasa},
  \citenamefont {Krumins}, \citenamefont {Gustafsson}, \citenamefont
  {Comparat}, \citenamefont {Auzins}, \citenamefont {Bergmann}, \citenamefont
  {Burian}, \citenamefont {Brusa}, \citenamefont {Castelli}, \citenamefont
  {Cerchiari}, \citenamefont {Ciuryło}, \citenamefont {Consolati},
  \citenamefont {Doser}, \citenamefont {Graczykowski}, \citenamefont
  {Grosbart}, \citenamefont {Guatieri}, \citenamefont {Haider}, \citenamefont
  {Janik}, \citenamefont {Kasprowicz}, \citenamefont {Khatri}, \citenamefont
  {Kłosowski}, \citenamefont {Kornakov}, \citenamefont {Lappo}, \citenamefont
  {Linek}, \citenamefont {Malamant}, \citenamefont {Mariazzi}, \citenamefont
  {Petracek}, \citenamefont {Piwiński}, \citenamefont {Pospíšil},
  \citenamefont {Povolo}, \citenamefont {Prelz}, \citenamefont {Rangwala},
  \citenamefont {Rauschendorfer}, \citenamefont {Rawat}, \citenamefont {Rodin},
  \citenamefont {Røhne}, \citenamefont {Sandaker}, \citenamefont
  {Smolyanskiy}, \citenamefont {Sowiński}, \citenamefont {Tefelski},
  \citenamefont {Vafeiadis}, \citenamefont {Welsch}, \citenamefont {Zawada},
  \citenamefont {Zielinski}, \citenamefont {Zurlo},\ and\ \citenamefont
  {{AEḡIS Collaboration}}}]{gloggler_positronium_2024}%
  \BibitemOpen
  \bibfield  {author} {\bibinfo {author} {\bibfnamefont {L.}~\bibnamefont
  {Glöggler}}, \bibinfo {author} {\bibfnamefont {N.}~\bibnamefont {Gusakova}},
  \bibinfo {author} {\bibfnamefont {B.}~\bibnamefont {Rienäcker}}, \bibinfo
  {author} {\bibfnamefont {A.}~\bibnamefont {Camper}}, \bibinfo {author}
  {\bibfnamefont {R.}~\bibnamefont {Caravita}}, \bibinfo {author}
  {\bibfnamefont {S.}~\bibnamefont {Huck}}, \bibinfo {author} {\bibfnamefont
  {M.}~\bibnamefont {Volponi}}, \bibinfo {author} {\bibfnamefont
  {T.}~\bibnamefont {Wolz}}, \bibinfo {author} {\bibfnamefont {L.}~\bibnamefont
  {Penasa}}, \bibinfo {author} {\bibfnamefont {V.}~\bibnamefont {Krumins}},
  \bibinfo {author} {\bibfnamefont {F.}~\bibnamefont {Gustafsson}}, \bibinfo
  {author} {\bibfnamefont {D.}~\bibnamefont {Comparat}}, \bibinfo {author}
  {\bibfnamefont {M.}~\bibnamefont {Auzins}}, \bibinfo {author} {\bibfnamefont
  {B.}~\bibnamefont {Bergmann}}, \bibinfo {author} {\bibfnamefont
  {P.}~\bibnamefont {Burian}}, \bibinfo {author} {\bibfnamefont
  {R.}~\bibnamefont {Brusa}}, \bibinfo {author} {\bibfnamefont
  {F.}~\bibnamefont {Castelli}}, \bibinfo {author} {\bibfnamefont
  {G.}~\bibnamefont {Cerchiari}}, \bibinfo {author} {\bibfnamefont
  {R.}~\bibnamefont {Ciuryło}}, \bibinfo {author} {\bibfnamefont
  {G.}~\bibnamefont {Consolati}}, \bibinfo {author} {\bibfnamefont
  {M.}~\bibnamefont {Doser}}, \bibinfo {author} {\bibfnamefont
  {{\L}.}~\bibnamefont {Graczykowski}}, \bibinfo {author} {\bibfnamefont
  {M.}~\bibnamefont {Grosbart}}, \bibinfo {author} {\bibfnamefont
  {F.}~\bibnamefont {Guatieri}}, \bibinfo {author} {\bibfnamefont
  {S.}~\bibnamefont {Haider}}, \bibinfo {author} {\bibfnamefont
  {M.}~\bibnamefont {Janik}}, \bibinfo {author} {\bibfnamefont
  {G.}~\bibnamefont {Kasprowicz}}, \bibinfo {author} {\bibfnamefont
  {G.}~\bibnamefont {Khatri}}, \bibinfo {author} {\bibfnamefont
  {{\L}.}~\bibnamefont {Kłosowski}}, \bibinfo {author} {\bibfnamefont
  {G.}~\bibnamefont {Kornakov}}, \bibinfo {author} {\bibfnamefont
  {L.}~\bibnamefont {Lappo}}, \bibinfo {author} {\bibfnamefont
  {A.}~\bibnamefont {Linek}}, \bibinfo {author} {\bibfnamefont
  {J.}~\bibnamefont {Malamant}}, \bibinfo {author} {\bibfnamefont
  {S.}~\bibnamefont {Mariazzi}}, \bibinfo {author} {\bibfnamefont
  {V.}~\bibnamefont {Petracek}}, \bibinfo {author} {\bibfnamefont
  {M.}~\bibnamefont {Piwiński}}, \bibinfo {author} {\bibfnamefont
  {S.}~\bibnamefont {Pospíšil}}, \bibinfo {author} {\bibfnamefont
  {L.}~\bibnamefont {Povolo}}, \bibinfo {author} {\bibfnamefont
  {F.}~\bibnamefont {Prelz}}, \bibinfo {author} {\bibfnamefont
  {S.}~\bibnamefont {Rangwala}}, \bibinfo {author} {\bibfnamefont
  {T.}~\bibnamefont {Rauschendorfer}}, \bibinfo {author} {\bibfnamefont
  {B.}~\bibnamefont {Rawat}}, \bibinfo {author} {\bibfnamefont
  {V.}~\bibnamefont {Rodin}}, \bibinfo {author} {\bibfnamefont
  {O.}~\bibnamefont {Røhne}}, \bibinfo {author} {\bibfnamefont
  {H.}~\bibnamefont {Sandaker}}, \bibinfo {author} {\bibfnamefont
  {P.}~\bibnamefont {Smolyanskiy}}, \bibinfo {author} {\bibfnamefont
  {T.}~\bibnamefont {Sowiński}}, \bibinfo {author} {\bibfnamefont
  {D.}~\bibnamefont {Tefelski}}, \bibinfo {author} {\bibfnamefont
  {T.}~\bibnamefont {Vafeiadis}}, \bibinfo {author} {\bibfnamefont
  {C.}~\bibnamefont {Welsch}}, \bibinfo {author} {\bibfnamefont
  {M.}~\bibnamefont {Zawada}}, \bibinfo {author} {\bibfnamefont
  {J.}~\bibnamefont {Zielinski}}, \bibinfo {author} {\bibfnamefont
  {N.}~\bibnamefont {Zurlo}},\ and\ \bibinfo {author} {\bibnamefont {{AEḡIS
  Collaboration}}},\ }\bibfield  {title} {\bibinfo {title} {Positronium {Laser}
  {Cooling} via the {1$^3S$-2$^3P$} {Transition} with a {Broadband} {Laser}
  {Pulse}},\ }\href {https://doi.org/10.1103/PhysRevLett.132.083402} {\bibfield
   {journal} {\bibinfo  {journal} {Phys. Rev. Lett.}\ }\textbf {\bibinfo
  {volume} {132}},\ \bibinfo {pages} {083402} (\bibinfo {year}
  {2024})}\BibitemShut {NoStop}%
\bibitem [{\citenamefont {Liang}\ and\ \citenamefont
  {Dermer}(1988)}]{liang_laser_1988}%
  \BibitemOpen
  \bibfield  {author} {\bibinfo {author} {\bibfnamefont {E.~P.}\ \bibnamefont
  {Liang}}\ and\ \bibinfo {author} {\bibfnamefont {C.~D.}\ \bibnamefont
  {Dermer}},\ }\bibfield  {title} {\bibinfo {title} {Laser cooling of
  positronium},\ }\href {https://doi.org/10.1016/0030-4018(88)90116-2}
  {\bibfield  {journal} {\bibinfo  {journal} {Opt. Commun.}\ }\textbf {\bibinfo
  {volume} {65}},\ \bibinfo {pages} {419} (\bibinfo {year} {1988})}\BibitemShut
  {NoStop}%
\bibitem [{\citenamefont {Kumita}\ \emph {et~al.}(2002)\citenamefont {Kumita},
  \citenamefont {Hirose}, \citenamefont {Irako}, \citenamefont {Kadoya},
  \citenamefont {Matsumoto}, \citenamefont {Wada}, \citenamefont {Mondal},
  \citenamefont {Yabu}, \citenamefont {Kobayashi},\ and\ \citenamefont
  {Kajita}}]{kumita_study_2002}%
  \BibitemOpen
  \bibfield  {author} {\bibinfo {author} {\bibfnamefont {T.}~\bibnamefont
  {Kumita}}, \bibinfo {author} {\bibfnamefont {T.}~\bibnamefont {Hirose}},
  \bibinfo {author} {\bibfnamefont {M.}~\bibnamefont {Irako}}, \bibinfo
  {author} {\bibfnamefont {K.}~\bibnamefont {Kadoya}}, \bibinfo {author}
  {\bibfnamefont {B.}~\bibnamefont {Matsumoto}}, \bibinfo {author}
  {\bibfnamefont {K.}~\bibnamefont {Wada}}, \bibinfo {author} {\bibfnamefont
  {N.}~\bibnamefont {Mondal}}, \bibinfo {author} {\bibfnamefont
  {H.}~\bibnamefont {Yabu}}, \bibinfo {author} {\bibfnamefont {K.}~\bibnamefont
  {Kobayashi}},\ and\ \bibinfo {author} {\bibfnamefont {M.}~\bibnamefont
  {Kajita}},\ }\bibfield  {title} {\bibinfo {title} {Study on laser cooling of
  ortho-positronium},\ }\href {https://doi.org/10.1016/S0168-583X(02)00863-7}
  {\bibfield  {journal} {\bibinfo  {journal} {Nucl. Instrum. Methods Phys. Res.
  B}\ }\textbf {\bibinfo {volume} {192}},\ \bibinfo {pages} {171} (\bibinfo
  {year} {2002})}\BibitemShut {NoStop}%
\bibitem [{\citenamefont {Shu}\ \emph {et~al.}(2016)\citenamefont {Shu},
  \citenamefont {Fan}, \citenamefont {Yamazaki}, \citenamefont {Namba},
  \citenamefont {Asai}, \citenamefont {Yoshioka},\ and\ \citenamefont
  {Kuwata-Gonokami}}]{shu_study_2016}%
  \BibitemOpen
  \bibfield  {author} {\bibinfo {author} {\bibfnamefont {K.}~\bibnamefont
  {Shu}}, \bibinfo {author} {\bibfnamefont {X.}~\bibnamefont {Fan}}, \bibinfo
  {author} {\bibfnamefont {T.}~\bibnamefont {Yamazaki}}, \bibinfo {author}
  {\bibfnamefont {T.}~\bibnamefont {Namba}}, \bibinfo {author} {\bibfnamefont
  {S.}~\bibnamefont {Asai}}, \bibinfo {author} {\bibfnamefont {K.}~\bibnamefont
  {Yoshioka}},\ and\ \bibinfo {author} {\bibfnamefont {M.}~\bibnamefont
  {Kuwata-Gonokami}},\ }\bibfield  {title} {\bibinfo {title} {Study on cooling
  of positronium for {Bose}–{Einstein} condensation},\ }\href
  {https://doi.org/10.1088/0953-4075/49/10/104001} {\bibfield  {journal}
  {\bibinfo  {journal} {J. Phys. B: At. Mol. Opt. Phys.}\ }\textbf {\bibinfo
  {volume} {49}},\ \bibinfo {pages} {104001} (\bibinfo {year}
  {2016})}\BibitemShut {NoStop}%
\bibitem [{\citenamefont {Zimmer}\ \emph {et~al.}(2021)\citenamefont {Zimmer},
  \citenamefont {Yzombard}, \citenamefont {Camper},\ and\ \citenamefont
  {Comparat}}]{zimmer_positronium_2021}%
  \BibitemOpen
  \bibfield  {author} {\bibinfo {author} {\bibfnamefont {C.}~\bibnamefont
  {Zimmer}}, \bibinfo {author} {\bibfnamefont {P.}~\bibnamefont {Yzombard}},
  \bibinfo {author} {\bibfnamefont {A.}~\bibnamefont {Camper}},\ and\ \bibinfo
  {author} {\bibfnamefont {D.}~\bibnamefont {Comparat}},\ }\bibfield  {title}
  {\bibinfo {title} {Positronium laser cooling in a magnetic field},\ }\href
  {https://doi.org/10.1103/PhysRevA.104.023106} {\bibfield  {journal} {\bibinfo
   {journal} {Phys. Rev. A}\ }\textbf {\bibinfo {volume} {104}},\ \bibinfo
  {pages} {023106} (\bibinfo {year} {2021})}\BibitemShut {NoStop}%
\bibitem [{\citenamefont {Yamada}\ \emph {et~al.}(2021)\citenamefont {Yamada},
  \citenamefont {Tajima}, \citenamefont {Murayoshi}, \citenamefont {Fan},
  \citenamefont {Ishida}, \citenamefont {Namba}, \citenamefont {Asai},
  \citenamefont {Kuwata-Gonokami}, \citenamefont {Chae}, \citenamefont {Shu},\
  and\ \citenamefont {Yoshioka}}]{yamada_theoretical_2021}%
  \BibitemOpen
  \bibfield  {author} {\bibinfo {author} {\bibfnamefont {K.}~\bibnamefont
  {Yamada}}, \bibinfo {author} {\bibfnamefont {Y.}~\bibnamefont {Tajima}},
  \bibinfo {author} {\bibfnamefont {T.}~\bibnamefont {Murayoshi}}, \bibinfo
  {author} {\bibfnamefont {X.}~\bibnamefont {Fan}}, \bibinfo {author}
  {\bibfnamefont {A.}~\bibnamefont {Ishida}}, \bibinfo {author} {\bibfnamefont
  {T.}~\bibnamefont {Namba}}, \bibinfo {author} {\bibfnamefont
  {S.}~\bibnamefont {Asai}}, \bibinfo {author} {\bibfnamefont {M.}~\bibnamefont
  {Kuwata-Gonokami}}, \bibinfo {author} {\bibfnamefont {E.}~\bibnamefont
  {Chae}}, \bibinfo {author} {\bibfnamefont {K.}~\bibnamefont {Shu}},\ and\
  \bibinfo {author} {\bibfnamefont {K.}~\bibnamefont {Yoshioka}},\ }\bibfield
  {title} {\bibinfo {title} {Theoretical {Analysis} and {Experimental}
  {Demonstration} of a {Chirped} {Pulse}-{Train} {Generator} and its
  {Potential} for {Efficient} {Cooling} of {Positronium}},\ }\href
  {https://doi.org/10.1103/PhysRevApplied.16.014009} {\bibfield  {journal}
  {\bibinfo  {journal} {Phys. Rev. Appl.}\ }\textbf {\bibinfo {volume} {16}},\
  \bibinfo {pages} {014009} (\bibinfo {year} {2021})}\BibitemShut {NoStop}%
\bibitem [{\citenamefont {Shu}\ \emph {et~al.}(2024{\natexlab{b}})\citenamefont
  {Shu}, \citenamefont {Miyamoto}, \citenamefont {Motohashi}, \citenamefont
  {Uozumi}, \citenamefont {Tajima},\ and\ \citenamefont
  {Yoshioka}}]{shu_development_2024}%
  \BibitemOpen
  \bibfield  {author} {\bibinfo {author} {\bibfnamefont {K.}~\bibnamefont
  {Shu}}, \bibinfo {author} {\bibfnamefont {N.}~\bibnamefont {Miyamoto}},
  \bibinfo {author} {\bibfnamefont {Y.}~\bibnamefont {Motohashi}}, \bibinfo
  {author} {\bibfnamefont {R.}~\bibnamefont {Uozumi}}, \bibinfo {author}
  {\bibfnamefont {Y.}~\bibnamefont {Tajima}},\ and\ \bibinfo {author}
  {\bibfnamefont {K.}~\bibnamefont {Yoshioka}},\ }\bibfield  {title} {\bibinfo
  {title} {Development of a laser for chirp cooling of positronium to near the
  recoil limit using a chirped pulse-train generator},\ }\href
  {https://doi.org/10.1103/PhysRevA.109.043520} {\bibfield  {journal} {\bibinfo
   {journal} {Phys. Rev. A}\ }\textbf {\bibinfo {volume} {109}},\ \bibinfo
  {pages} {043520} (\bibinfo {year} {2024}{\natexlab{b}})}\BibitemShut
  {NoStop}%
\bibitem [{\citenamefont {Meystre}\ and\ \citenamefont
  {Sargent}(2007)}]{meystre2007elements}%
  \BibitemOpen
  \bibfield  {author} {\bibinfo {author} {\bibfnamefont {P.}~\bibnamefont
  {Meystre}}\ and\ \bibinfo {author} {\bibfnamefont {M.}~\bibnamefont
  {Sargent}},\ }\href {https://doi.org/10.1007/978-3-540-74211-8} {\emph
  {\bibinfo {title} {Elements of Quantum Optics}}},\ \bibinfo {edition} {4th}\
  ed.\ (\bibinfo  {publisher} {Springer},\ \bibinfo {address} {Berlin,
  Heidelberg},\ \bibinfo {year} {2007})\BibitemShut {NoStop}%
\bibitem [{\citenamefont {Ilinova}\ \emph {et~al.}(2011)\citenamefont
  {Ilinova}, \citenamefont {Ahmad},\ and\ \citenamefont
  {Derevianko}}]{ilinova_doppler_2011}%
  \BibitemOpen
  \bibfield  {author} {\bibinfo {author} {\bibfnamefont {E.}~\bibnamefont
  {Ilinova}}, \bibinfo {author} {\bibfnamefont {M.}~\bibnamefont {Ahmad}},\
  and\ \bibinfo {author} {\bibfnamefont {A.}~\bibnamefont {Derevianko}},\
  }\bibfield  {title} {\bibinfo {title} {Doppler cooling with coherent trains
  of laser pulses and a tunable velocity comb},\ }\href
  {https://doi.org/10.1103/PhysRevA.84.033421} {\bibfield  {journal} {\bibinfo
  {journal} {Phys. Rev. A}\ }\textbf {\bibinfo {volume} {84}},\ \bibinfo
  {pages} {033421} (\bibinfo {year} {2011})}\BibitemShut {NoStop}%
\bibitem [{\citenamefont {Aspect}\ \emph {et~al.}(1989)\citenamefont {Aspect},
  \citenamefont {Arimondo}, \citenamefont {Kaiser}, \citenamefont
  {Vansteenkiste},\ and\ \citenamefont {Cohen-Tannoudji}}]{aspect_laser_1989}%
  \BibitemOpen
  \bibfield  {author} {\bibinfo {author} {\bibfnamefont {A.}~\bibnamefont
  {Aspect}}, \bibinfo {author} {\bibfnamefont {E.}~\bibnamefont {Arimondo}},
  \bibinfo {author} {\bibfnamefont {R.}~\bibnamefont {Kaiser}}, \bibinfo
  {author} {\bibfnamefont {N.}~\bibnamefont {Vansteenkiste}},\ and\ \bibinfo
  {author} {\bibfnamefont {C.}~\bibnamefont {Cohen-Tannoudji}},\ }\bibfield
  {title} {\bibinfo {title} {Laser cooling below the one-photon recoil energy
  by velocity-selective coherent population trapping: theoretical analysis},\
  }\href {https://doi.org/10.1364/JOSAB.6.002112} {\bibfield  {journal}
  {\bibinfo  {journal} {J. Opt. Soc. Am. B}\ }\textbf {\bibinfo {volume} {6}},\
  \bibinfo {pages} {2112} (\bibinfo {year} {1989})}\BibitemShut {NoStop}%
\bibitem [{\citenamefont {Papoff}\ \emph {et~al.}(1992)\citenamefont {Papoff},
  \citenamefont {Mauri},\ and\ \citenamefont
  {Arimondo}}]{papoff_transient_1992}%
  \BibitemOpen
  \bibfield  {author} {\bibinfo {author} {\bibfnamefont {F.}~\bibnamefont
  {Papoff}}, \bibinfo {author} {\bibfnamefont {F.}~\bibnamefont {Mauri}},\ and\
  \bibinfo {author} {\bibfnamefont {E.}~\bibnamefont {Arimondo}},\ }\bibfield
  {title} {\bibinfo {title} {Transient velocity-selective coherent population
  trapping in one dimension},\ }\href {https://doi.org/10.1364/JOSAB.9.000321}
  {\bibfield  {journal} {\bibinfo  {journal} {J. Opt. Soc. Am. B}\ }\textbf
  {\bibinfo {volume} {9}},\ \bibinfo {pages} {321} (\bibinfo {year}
  {1992})}\BibitemShut {NoStop}%
\bibitem [{\citenamefont {Malamant}\ \emph {et~al.}(2024)\citenamefont
  {Malamant}, \citenamefont {Gusakova}, \citenamefont {Sandaker}, \citenamefont
  {Sorokina}, \citenamefont {Comparat},\ and\ \citenamefont
  {Camper}}]{malamant_coherent_2024}%
  \BibitemOpen
  \bibfield  {author} {\bibinfo {author} {\bibfnamefont {J.}~\bibnamefont
  {Malamant}}, \bibinfo {author} {\bibfnamefont {N.}~\bibnamefont {Gusakova}},
  \bibinfo {author} {\bibfnamefont {H.}~\bibnamefont {Sandaker}}, \bibinfo
  {author} {\bibfnamefont {I.~T.}\ \bibnamefont {Sorokina}}, \bibinfo {author}
  {\bibfnamefont {D.}~\bibnamefont {Comparat}},\ and\ \bibinfo {author}
  {\bibfnamefont {A.}~\bibnamefont {Camper}},\ }\bibfield  {title} {\bibinfo
  {title} {Coherent laser cooling with trains of ultrashort laser pulses},\
  }\href {https://doi.org/10.1103/PhysRevA.110.013109} {\bibfield  {journal}
  {\bibinfo  {journal} {Phys. Rev. A}\ }\textbf {\bibinfo {volume} {110}},\
  \bibinfo {pages} {013109} (\bibinfo {year} {2024})}\BibitemShut {NoStop}%
\end{thebibliography}
\end{document}